\documentstyle[12pt,epsf]{article}
\newlength{\dinwidth}                                                          
\newlength{\dinmargin}                                                         
\begin{document}
\vspace{1 cm}
\newcommand{\Gev}       {\mbox{${\rm GeV}$}}
\newcommand{\Gevsq}     {\mbox{${\rm GeV}^2$}}
\newcommand{\qsd}       {\mbox{${Q^2}$}} 
\newcommand{\x}         {\mbox{${\it x}$}}
\newcommand{\smallqsd}  {\mbox{${q^2}$}} 
\newcommand{\ra}        {\mbox{$ \rightarrow $}}
\newcommand {\pom}  {I\hspace{-0.2em}P}
\newcommand {\alphapom} {\mbox{$\alpha_{_{\pom}}$}}
\newcommand {\xpom} {\mbox{$x_{_{\pom}}$}}
\newcommand {\xpomp}[1] {\mbox{$x^{#1}_{_{\pom}}\;$}}
\newcommand {\xpoma} {\mbox{$(1/x_{_{\pom}})^a\;$}}
% ---- commands from paul -----
\def\ctr#1{{\it #1}\\\vspace{10pt}}
\def\si{{\rm si}}
\def\Si{{\rm Si}}
\def\Ci{{\rm Ci}}
\def\qsq{Q^{2}}
\def\yjb{y_{_{JB}}}
\def\xjb{x_{_{JB}}}
\def\qjb{\qsq_{_{JB}}}
\def\gap{\hspace{0.5cm}}
\renewcommand{\thefootnote}{\arabic{footnote}}
\title {
\hfill{\normalsize  DESY 98-140}\\
\hfill{\normalsize  8 September 1998}\\
\hfill{\normalsize   Rev. 4 January 1999}\\
\vspace{0.5cm}
{\bf Enhancing Squark/Leptoquark Production by Increasing the HERA Beam 
Energies
 }\\ 
\author{
Michiel Botje \\
{NIKHEF}\\
G\"{u}nter Wolf\\
{Deutsches Elektronen Synchrotron DESY  }\\
 }}
\date{}
\maketitle
\begin{abstract}

The cross section for production of R-parity violating squarks ($\tilde{q}$) or leptoquarks ($LQ$) with masses above 200~GeV
by positron-proton scattering at HERA can be substantially increased by a 
moderate increase of the beam energies. 
If, for example, the positron energy is raised from 27.5 to 32.5~GeV and the
proton energy from 820 to 1000~GeV, the $\tilde{q}/LQ$ production
cross section for masses of 200 -- 230~GeV increases by a factor of about 2--6 and the search range for $\tilde{q}/LQ$ is extended by about 40~GeV to masses of 270 -- 280~GeV. In this report event yields are presented for beam energies
in the range $E_e =$ 27.5--35~GeV  and $E_p =$ 820--1100~GeV. Also presented
is the rate of background events expected from Standard Model neutral and charged current
$e^{\pm}p$ scattering. 
\end{abstract}
\setcounter{page}{1}
\thispagestyle{empty}   
%\newpage
%
%
\section{\bf Introduction}
\label{s:intro}
\begin{figure}[p]
\includegraphics{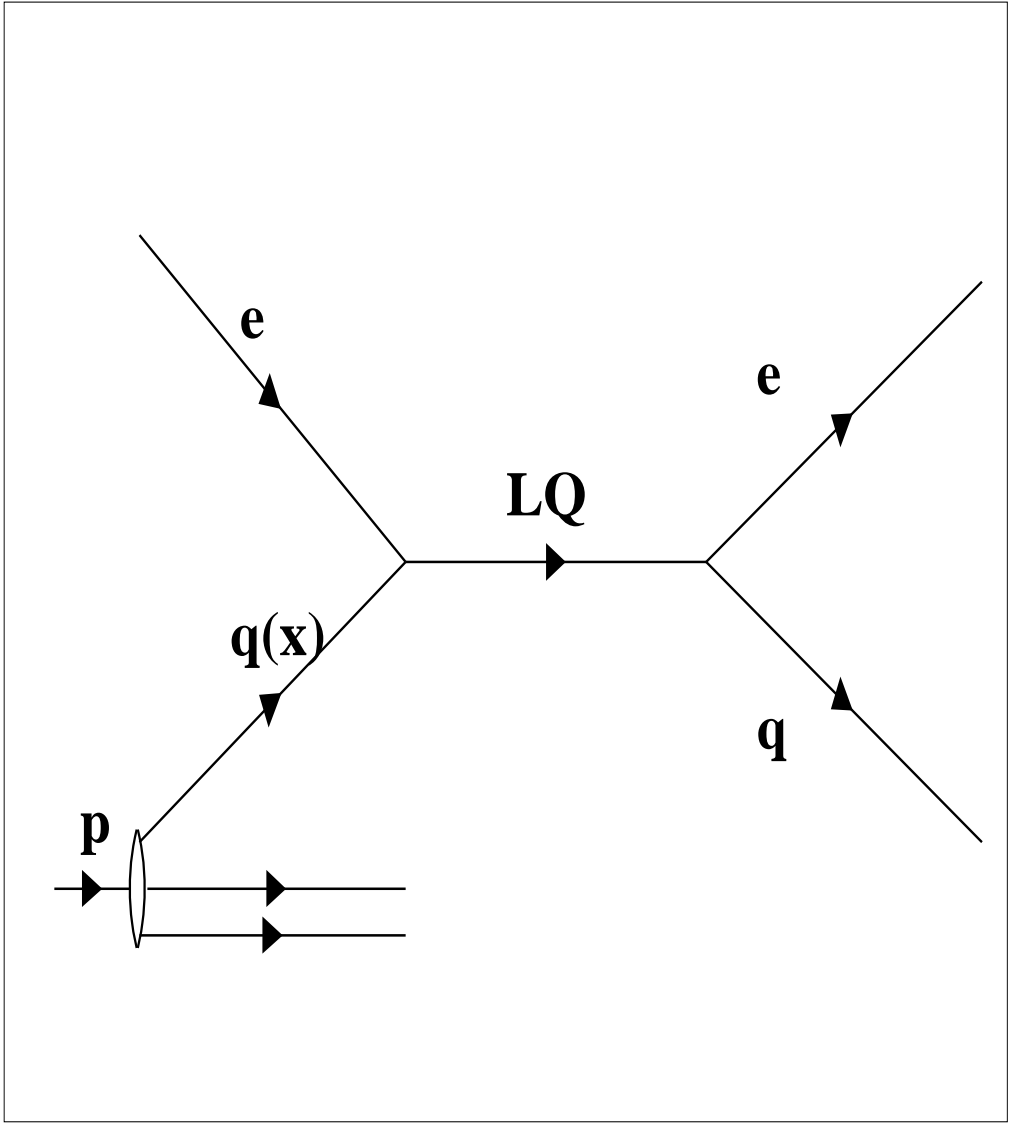}
\unitlength1cm
\begin{picture}(15,18)
\thicklines
\end{picture}
\caption{Diagram for leptoquark production by $ep$ scattering.}
\label{f:lqdiag}
\end{figure}

The ZEUS~\cite{Zhixhiy} and H1~\cite{H1hixhiy} experiments, studying 
deep-inelastic positron proton scattering (DIS) at HERA, recently reported 
an excess of events in the region of high Bjorken-$x$ and large negative 
photon virtualities ($Q^2$)~\footnote{Preliminary results obtained recently after more than doubling the integrated luminosity did not show an increase of the excess, see~\cite{Zhixhiy98,H1hixhiy98}.}. This excess may result from the production of a R-parity violating squark $\tilde{q}$ or of a  
leptoquark $LQ$  via the fusion of a positron  and a quark 
(see Fig.~\ref{f:lqdiag}), 
\begin{eqnarray}
e^+ p\,  \to \, \tilde{q}/LQ \, X\, \to e^{\prime} + h + X\, .
\label{eq:leptoquark}
\end{eqnarray}

For the purpose of this note, $\tilde{q}$ and $LQ$ are interchangeable. The symbol $LQ$ will be used to denote either one of them.
The data indicate for the mass of the $LQ$ a value around $M_{LQ} =$
200--230~GeV which corresponds to a proton momentum fraction $x$ 
carried by the fusing 
quark of $x = M^2_{LQ}/s =$ 0.45--0.60, where $s = 4E_eE_p$ is the square of 
the total $ep$ c.m. energy and $E_e,E_p$ are the positron and proton beam
energies. The possible existence of $\tilde{q}$ or $LQ$ in this mass range prompted us to study the $\tilde{q}/LQ$ production rate as a function of the HERA beam energies.

 The lowest order cross section for producing a scalar R-parity violating squark $\tilde{q}$ or leptoquark $LQ$
in $ep$ collisions can be written as (see e.g. ~\cite{Rueckl})
\begin{eqnarray} \label{e:lqxsec}
\frac{d\sigma_{ep \to LQ}}{dy} = \frac{\pi}{4s} \lambda^2 q(x,Q^2) = 
\frac{\pi}{4M^2_{LQ}} \lambda^2 x q(x,Q^2)
\label{lqcross}
\end{eqnarray}
where $Q^2 = - (e - e^{\prime})^2$ and $y = Q^2/(x s)$ 
are the standard DIS variables and $x q(x,Q^2)$ is the momentum density of 
the quark of type $q$ in the proton fusing with an electron or positron, generically denoted by $e$; $\lambda$ is a coupling constant which depends on the type of quark and lepton that form the $\tilde{q}/LQ$ state. The large value of $x$ plus the small production cross section combined with the absence of a signal in the electron-proton data favors the assumption that the fusing quark 
is an up or down quark (see e.g.~\cite{Wilczek}). Since the density of anti-sea quarks is small at large $x$ we expect that the $LQ$ production cross section is proportional to the momentum densities of the quarks, $x u_v(x,Q^2)$, $x d_v(x,Q^2)$ or $x s(x,Q^2)$, which are steeply falling with $x$ at high $x$. For this reason, a large increase of the production cross section can be obtained by a moderate increase of the HERA beam energies.

This note presents the expected increase of the $\tilde{q}/LQ$ production cross section and the $\tilde{q}/LQ$ search range for various combinations of positron and proton beam energies. Also given are the number of background events predicted by the Standard Model (SM) for neutral current (NC) and charged current (CC) $e^{\pm}p$ scattering. The data reported in~\cite{Zhixhiy,H1hixhiy} were taken at the ('nominal') beam energies of
$E_e = 27.5$~GeV, $E_p = 820$~GeV.  The range explored in this study is $E_e =$ 27.5--35~GeV and $E_p =$ 820--1100~GeV. The maximum values of $E_e = 35$~GeV and $E_p = 1100$~GeV should be considered as extreme upper limits for HERA. 

\section{\bf Choice of parameters}
\label{s:parameters}

The scalar $LQ$ (or $\tilde{q}$) production cross sections were calculated from
eq.~(\ref{e:lqxsec}) for the processes $e^+d \rightarrow LQ$,
$e^+u \rightarrow LQ$ and $e^+s \rightarrow LQ$. 
The quark momentum densities were
taken from the parton distribution set MRSA~\cite{mrsa}. In order to fix
the production rate of $LQ$ events the $LQ$ cross section was integrated 
over $y > 0.25$ and normalized such that the number of $LQ$ candidates 
accepted in $e^+p$ NC scattering as reported in ref.~\cite{Zhixhiy,H1hixhiy,Zhixhiy98,H1hixhiy98} is approximately reproduced. Taking $M_{LQ} = 215$ GeV and assuming 10 $LQ$ events for an integrated luminosity of  $L = 50$~pb$^{-1}$  at the nominal beam energies $E_e$ ($E_p$) = 27.5 (820)~GeV yields for the coupling constant $\lambda$ the values 0.045, 0.021, 0.7 for the $e^+d$, $e^+u$ and $e^+s$ states, respectively. For simplicity, we assume that the acceptance for $LQ$ candidates does not depend on whether a NC or a CC event sample is selected.

The background from the standard NC processes,
$e^{\pm}p \rightarrow e^{\pm}X$, was estimated by calculating the differential cross sections
\begin{equation} \label{ncxsec}
\frac{d^2 \sigma_{NC}^{\pm}}{dxdy} = 
\frac{2 \pi \alpha^2 s Y_+}{Q^4} \left\{
F_{2,NC} - \frac{y^2}{Y_+} F_{L,NC}
\mp \frac{Y_-}{Y_+}xF_{3,NC} \right\},
\end{equation}
taking into account the contributions
from $\gamma Z$ interference and $Z$ exchange~\cite{ingelman}.
Likewise the cross sections
\begin{equation} \label{ccxsec}
\frac{d^2 \sigma_{CC}^{\pm}}{dxdy} = 
\frac{G^2_F s Y_+}{8 \pi} \frac{M_W^4}{(Q^2 + M_W^2)^2} \left\{
F_{2,CC}^{\pm} - \frac{y^2}{Y_+} F_{L,CC}^{\pm}
\mp \frac{Y_-}{Y_+}xF_{3,CC}^{\pm} \right\}
\end{equation}
were calculated for the CC processes
$e^+p \rightarrow \bar{\nu}X$ and
$e^-p \rightarrow \nu X$.
In eqs.~(\ref{ncxsec}) and~(\ref{ccxsec}) $\alpha$ is the fine structure
constant, $G_F$ the Fermi constant, $M_W$ the mass of the $W$ and
$Y_{\pm} = 1 \pm (1-y)^2$.
The structure functions $F_2$ and $xF_3$ were calculated
in leading order with the parton distributions
of the proton taken from MRSA. The contribution from $F_L$ in this region of high $x$ is expected to be small and was neglected. The cross sections were integrated over $y > 0.25$ and over the band $x = M^2_{LQ}/s \pm 0.08$ which corresponds roughly to $\pm$ 2 standard deviations for the $LQ$ mass measurement of~\cite{Zhixhiy,H1hixhiy}. The number of background events was obtained by normalizing the $e^+p$ NC cross section to a yield of 0.675 accepted events in the bin $0.55 < x < 0.65$, $y > 0.25$ and $L = 20$~pb$^{-1}$~\cite{Zhixhiy}. 
In this way the acceptance quoted in ~\cite{Zhixhiy} for $e^+p$ NC 
scattering is approximately taken into account. 
The same acceptance was assumed for $e^-p$ NC and $e^{\pm}p$ CC scattering.

\section{Signal and background for $M_{LQ}$ = 200 - 230 GeV}

Tables~\ref{t:tab1} and~\ref{t:tab2} give the estimated event
yields ($N^i_{LQ}$) for $e^+q_i \rightarrow LQ$, $q_i = d,u,s$ with
an integrated luminosity of $L = 50$~pb$^{-1}$ for
various combinations of $E_e =$ 27.5--35~GeV, $E_p =$ 820--1100~GeV
and $M_{LQ}$ = 200--230~GeV. 
Also given are the expected number of $e^+p$ and $e^-p$ NC and CC
background events ($N_{NC,CC}^+$, $N_{NC,CC}^-$).
It is seen that for $e^+p$ scattering the CC background is about a factor of
five lower than the NC background whereas in $e^-p$ scattering the CC and NC
contributions are roughly equal in size. The total background in $e^+p$
scattering is about a factor of four lower than in $e^-p$.

Tables~\ref{t:tab3} and~\ref{t:tab4} present the gain in $LQ$ production,
defined as
\[
G(E_e,E_p,M_{LQ}) = \frac{N_{LQ}(E_e,E_p,M_{LQ})}{N_{LQ}(E_e^0,E_p^0,M_{LQ})}
\]
where $E_e^0$ $(E_p^0) = 27.5$ (820)~GeV denote the nominal beam energies.
Also listed are the signal over background ratios
\[
S^+_{CC,NC} = \frac{N_{LQ}}{N^+_{CC,NC}}
\]
for $e^+p$ scattering.

\begin{figure}[p]
\includegraphics{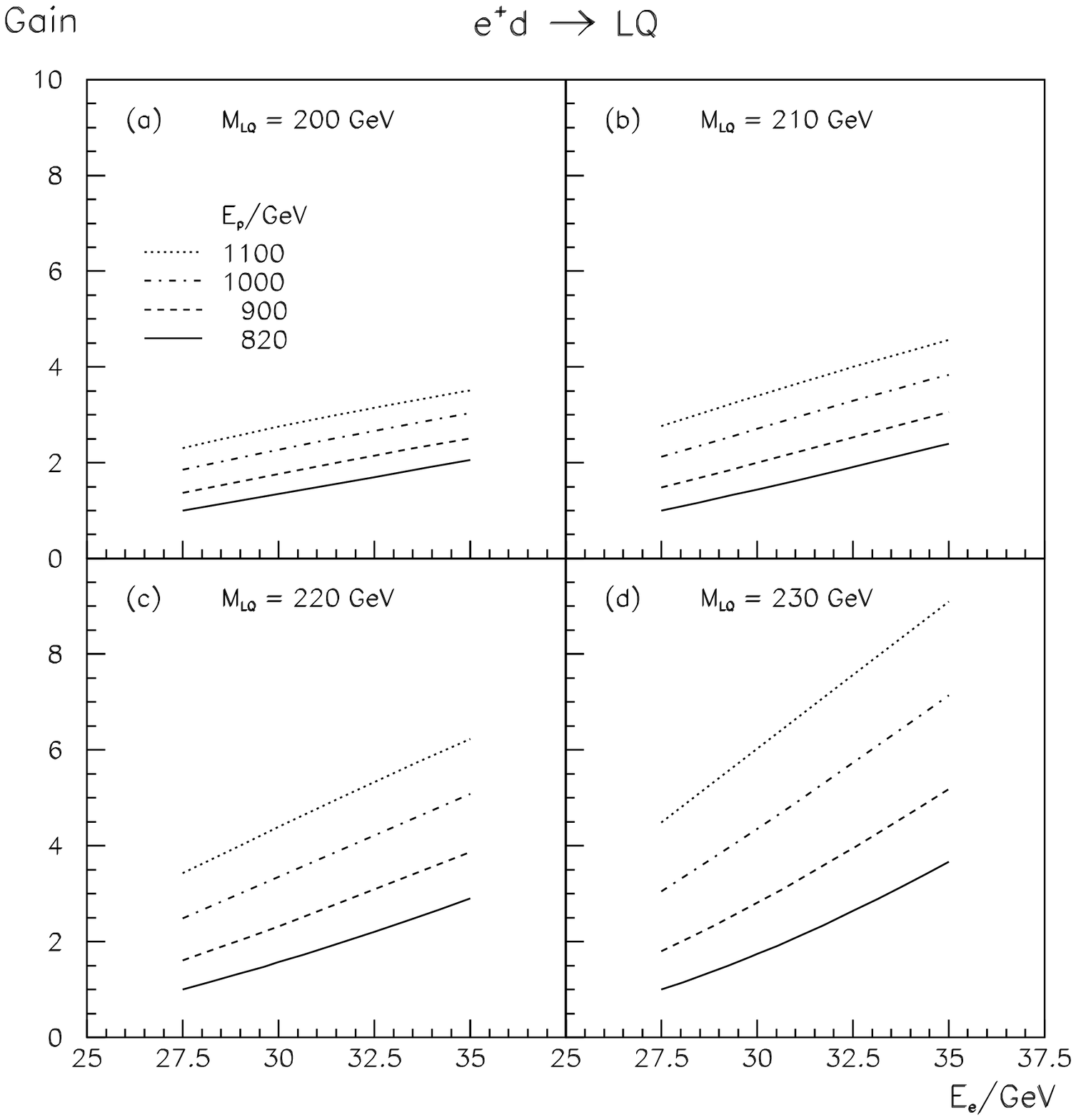}
\unitlength1cm
\begin{picture}(15,18)
\thicklines
\end{picture}
\caption{
The gain $G = {N_{LQ}(E_e,E_p)}/{N_{LQ}(27.5,820)}$ 
in squark/leptoquark production by $e^+d$ scattering as function of the
positron energy $E_e$ for various proton energies $E_p$ and assumed squark/leptoquark
masses of (a)~$M_{LQ} = 200$~GeV; (b)~210~GeV; (c)~220~GeV and (d)~230~GeV.
}
\protect\label{f:gaind}
\end{figure}

\begin{figure}[p]
\includegraphics{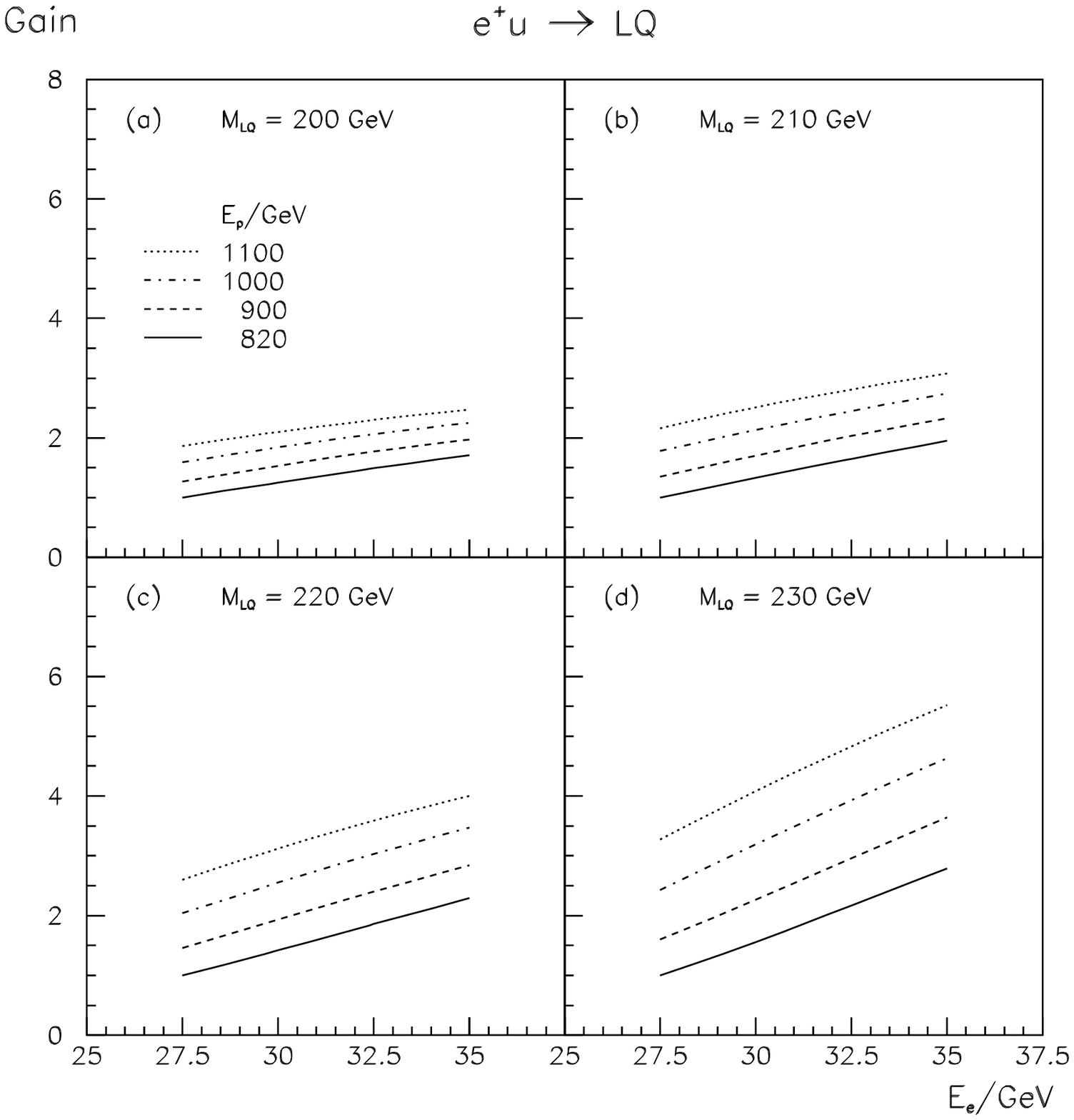}
\unitlength1cm
\begin{picture}(15,18)
\thicklines
\end{picture}
\caption{
The gain $G = {N_{LQ}(E_e,E_p)}/{N_{LQ}(27.5,820)}$ 
in squark/leptoquark production by $e^+u$ scattering as function of the
positron energy $E_e$ for various proton energies $E_p$ and assumed squark/leptoquark
masses of (a)~$M_{LQ} = 200$~GeV; (b)~210~GeV; (c)~220~GeV and (d)~230~GeV.
}
\protect\label{f:gainu}
\end{figure}

\begin{figure}[p]
\includegraphics{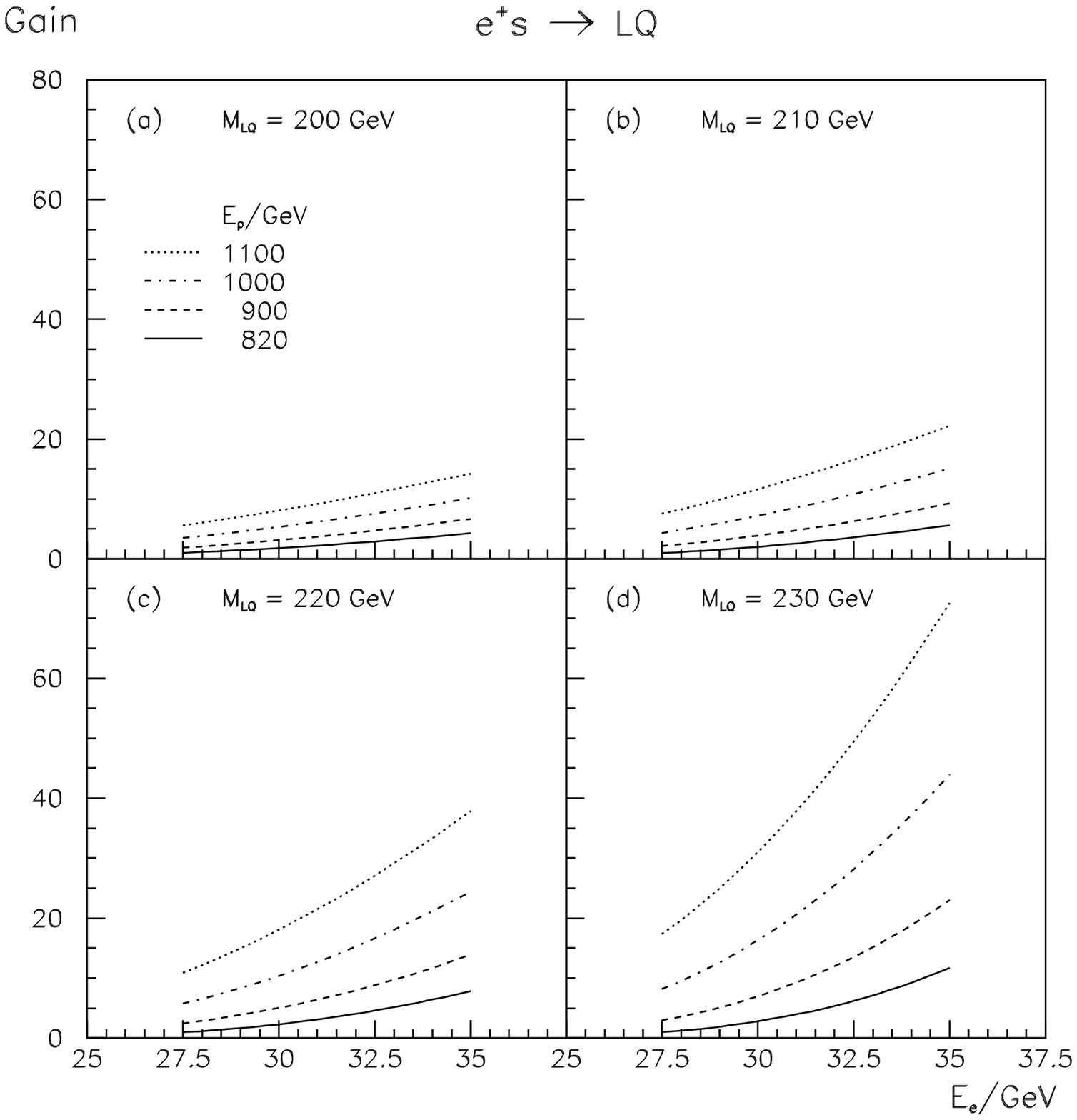}
\unitlength1cm
\begin{picture}(15,18)
\thicklines
\end{picture}
\caption{
The gain $G = {N_{LQ}(E_e,E_p)}/{N_{LQ}(27.5,820)}$ 
in squark/leptoquark production by $e^+s$ scattering as function of the
positron energy $E_e$ for various proton energies $E_p$ and assumed squark/leptoquark
masses of (a)~$M_{LQ} = 200$~GeV; (b)~210~GeV; (c)~220~GeV and (d)~230~GeV.
}
\protect\label{f:gains}
\end{figure}

\begin{figure}[p]
\includegraphics{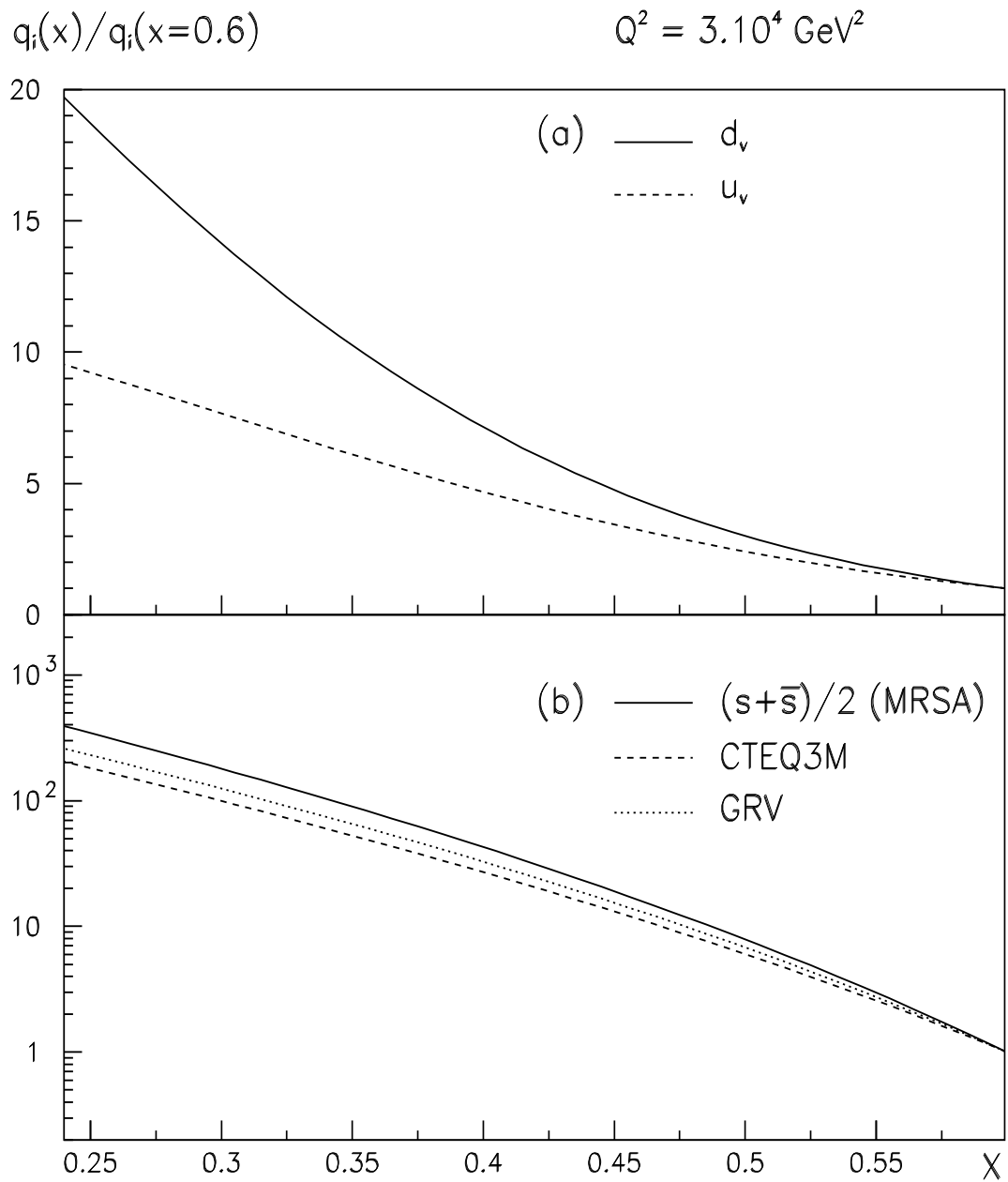}
\unitlength1cm
\begin{picture}(15,18)
\thicklines
\end{picture}
\caption{
(a) The down (solid curve) and up (dashed curve) valence distributions 
and (b) the strange quark distribution versus
$x$ at a fixed value of $Q^2 = 3 \times 10^4$ GeV$^2$. The distributions
are taken from the parton distribution set MRSA and are normalized
to unity at $x = 0.6$. Also shown are the strange quark distributions
from the sets CTEQ3M and GRV.
}
\protect\label{f:dus3e4}
\end{figure}

In Figs.~\ref{f:gaind} and~\ref{f:gainu}
we show the gain in $LQ$ events
with respect to production at
nominal beam energies for $e^+d \to LQ$ and $e^+u \to LQ$,
respectively. The gain is larger for larger leptoquark masses mainly
because the shift towards smaller $x_{LQ}$ is larger when $M_{LQ}$
increases, see Tables~\ref{t:tab1} and~\ref{t:tab2}.
Comparing Figs.~\ref{f:gaind} and~\ref{f:gainu} it is seen that a larger
gain is obtained for the $e^+d$ as compared to the $e^+u$ case. This is so because the
$d_v$ distribution is considerably steeper at high $x$ than the 
$u_v$ distribution as illustrated in Fig.~\ref{f:dus3e4}a.
For $e^+s \to LQ$ production the gain is particularly large (see Fig.~\ref{f:gains}) because the strange quark distribution is a very steep
function of $x$ at large $x$, as shown in Fig.~\ref{f:dus3e4}b.
However, the strange quark distribution in the proton is not very well
known. This is illustrated in Fig.~\ref{f:dus3e4}b, where
the distribution from MRSA is compared
with those from CTEQ3M~\cite{cteq3m} and GRV~\cite{grv}: the gain in
$LQ$ production from $s$ quarks is about a factor of 1.5 lower if
CTEQ3M instead of MRSA parton distributions are used.

\begin{figure}[p]
\includegraphics{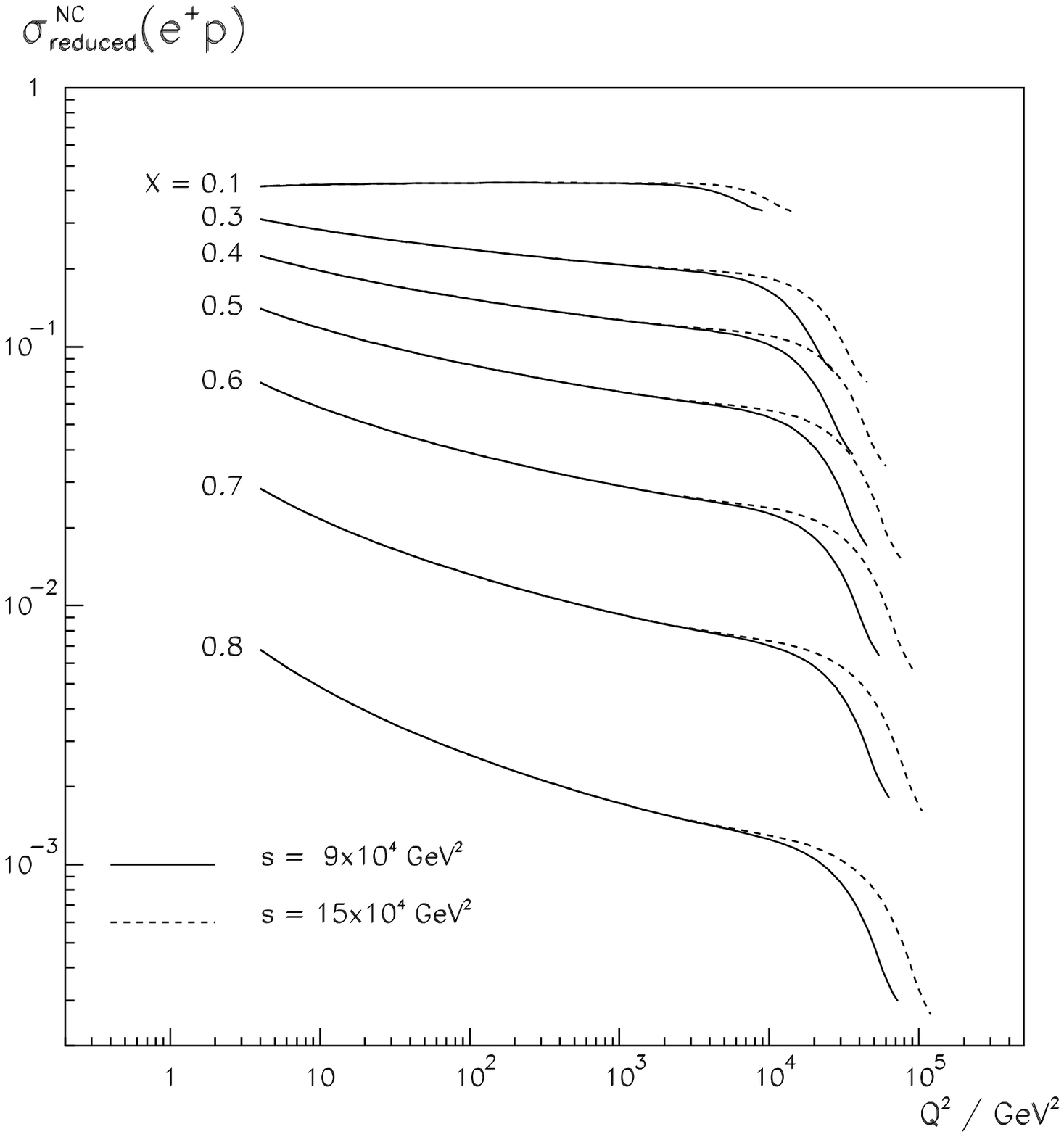}
\unitlength1cm
\begin{picture}(15,18)
\thicklines
\end{picture}
\vspace{1cm}
\caption{The reduced cross section (see text) for the NC process
$e^+p \to e^+X$.
}
\label{f:eplusnc}
\end{figure}

\begin{figure}[p]
\includegraphics{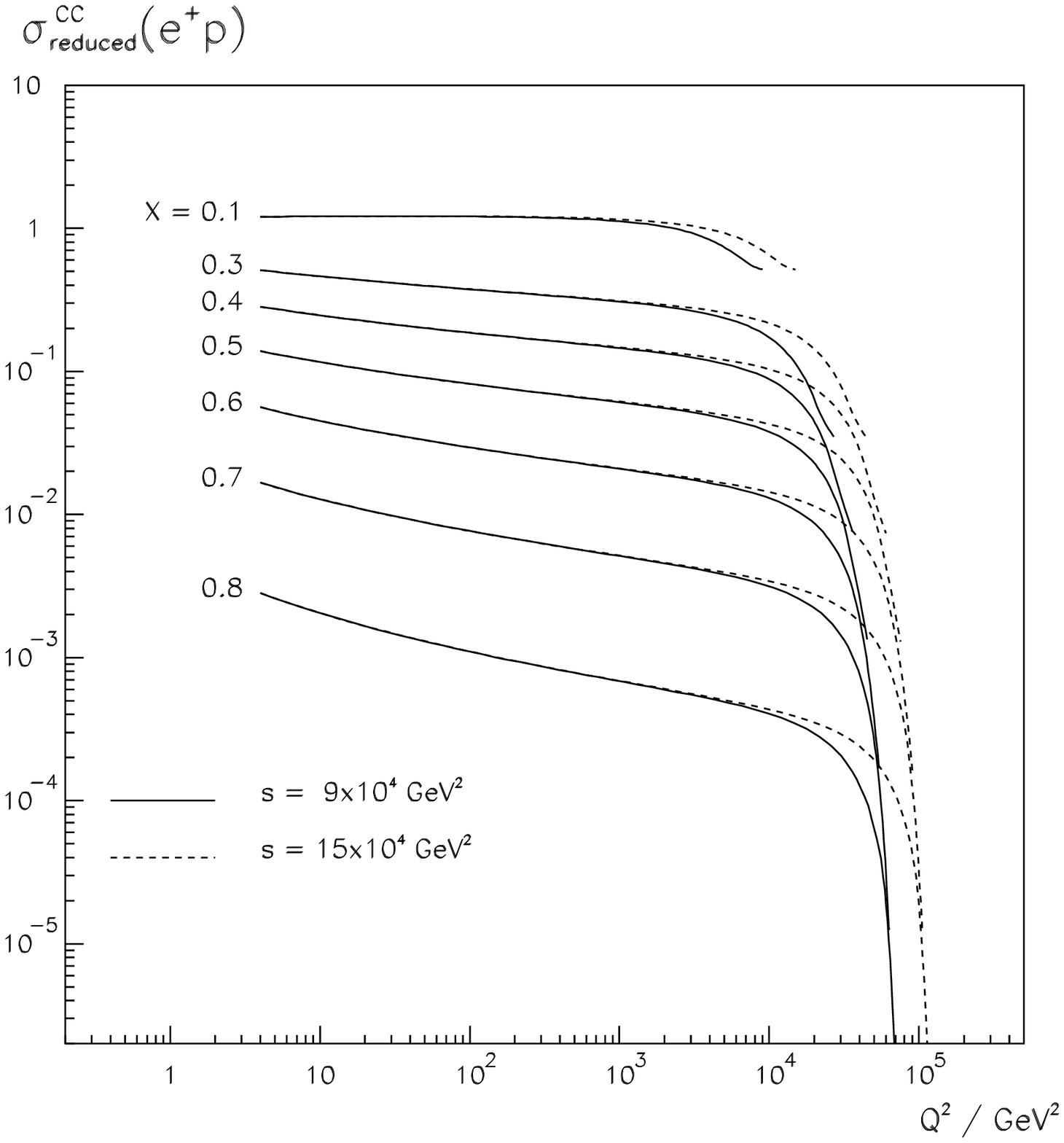}
\unitlength1cm
\begin{picture}(15,18)
\thicklines
\end{picture}
\vspace{1cm}
\caption{The reduced cross section (see text) for the CC process
$e^+p \to \bar{\nu}X$.
}
\label{f:epluscc}
\end{figure}

\begin{figure}[p]
\includegraphics{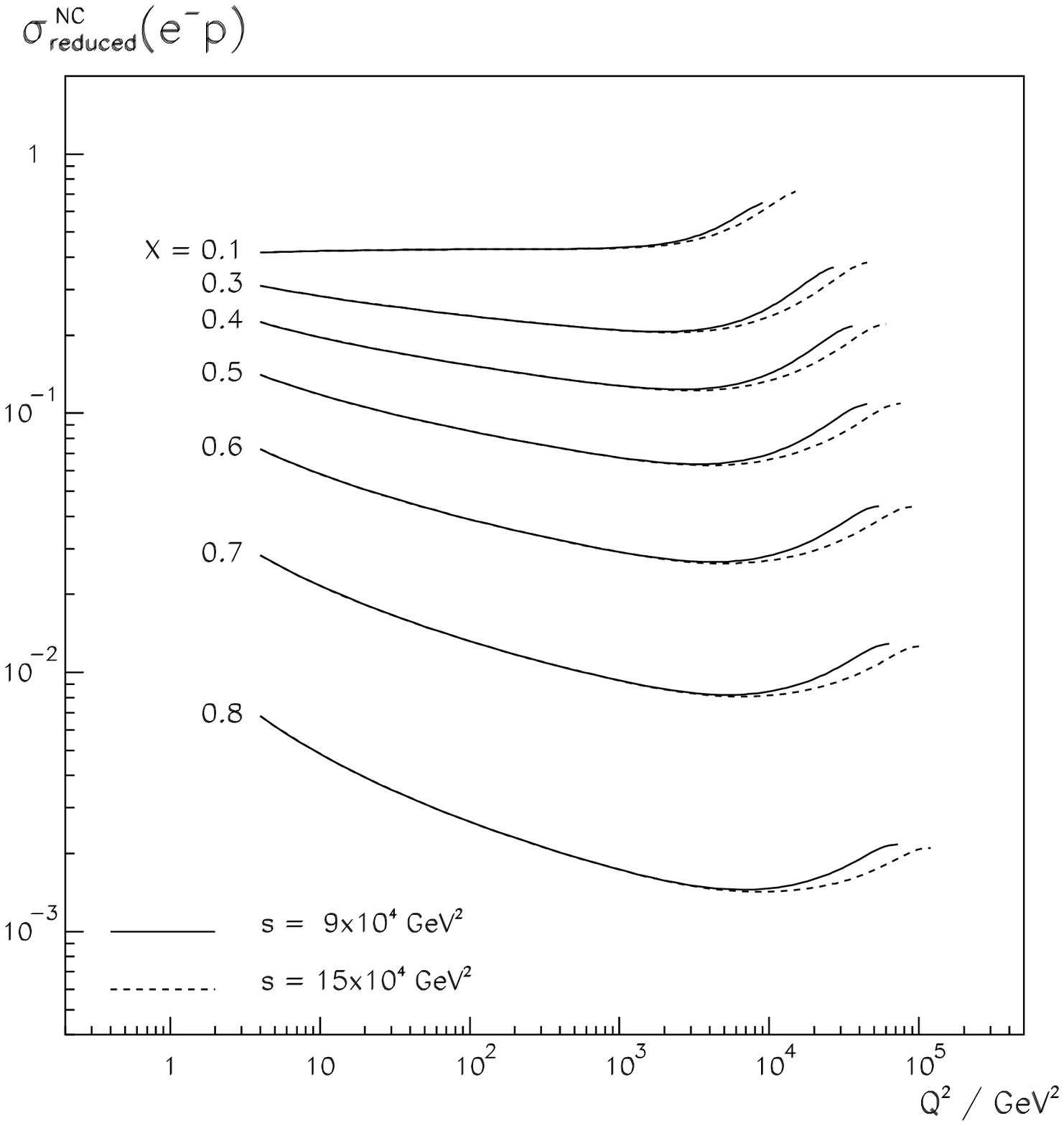}
\unitlength1cm
\begin{picture}(15,18)
\thicklines
\end{picture}
\vspace{1cm}
\caption{The reduced cross section (see text) for the NC process
$e^-p \to e^-X$.
}
\label{f:eminnc}
\end{figure}

\begin{figure}[p]
\includegraphics{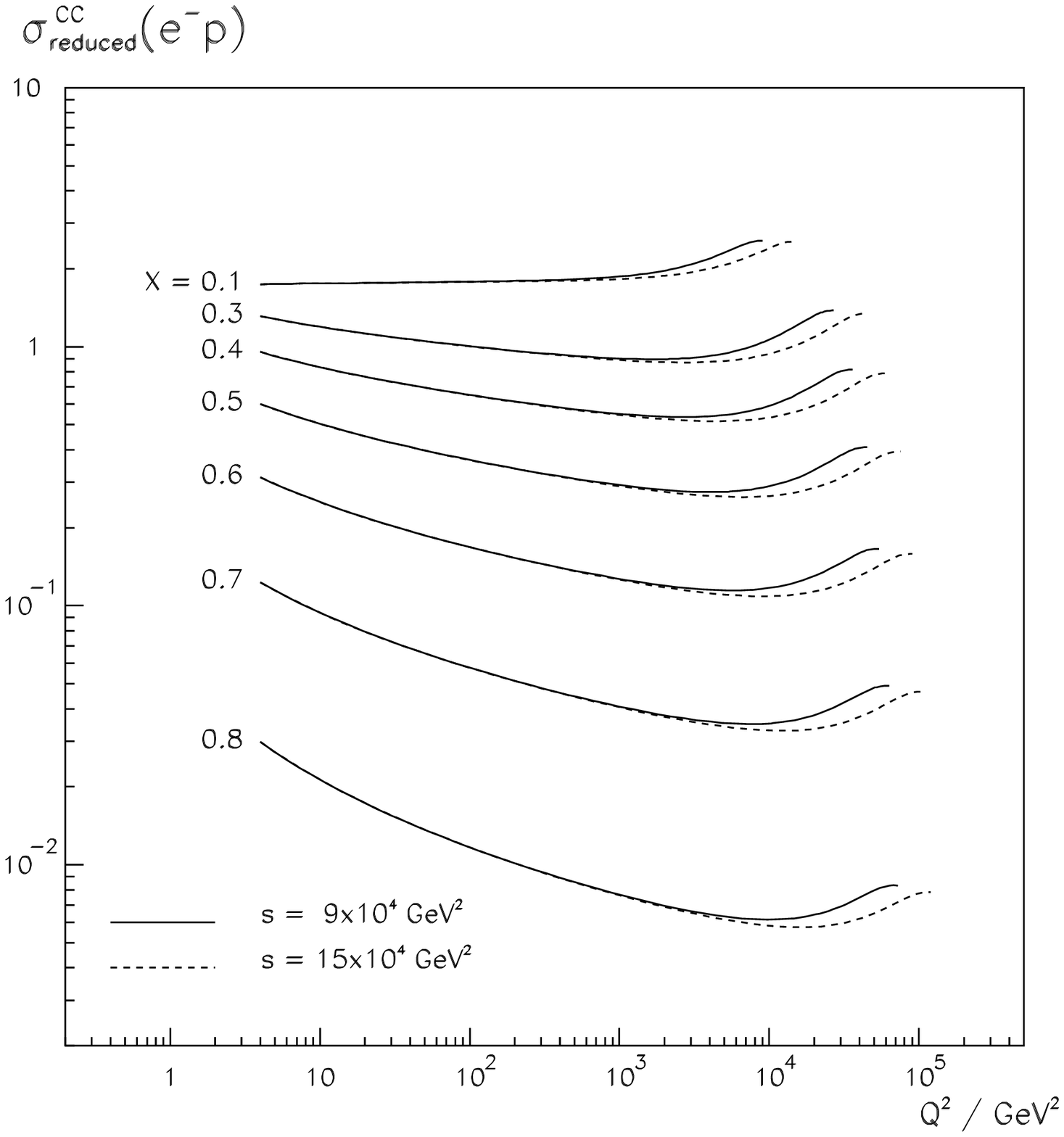}
\unitlength1cm
\begin{picture}(15,18)
\thicklines
\end{picture}
\vspace{1cm}
\caption{The reduced cross section (see text) for the CC process
$e^-p \to \nu X$.
}
\label{f:emincc}
\end{figure}

Figures~\ref{f:eplusnc}--\ref{f:emincc} show the $e^+$ and
$e^-$ NC and CC `reduced' cross
sections defined as the terms inside the curly brackets in
eqs.~(\ref{ncxsec}) and~(\ref{ccxsec}) for $s = 9\times 10^4$ GeV$^2$ (corresponding to the nominal beam energies) and $s = 15\times 10^4$ GeV$^2$ ($E_e = 35$ GeV, $E_p = 1100$ GeV). For $e^+p$ scattering, the reduced NC and CC cross sections at fixed $Q^2$ above the masses squared of W, Z increase with $s$ while for $e^- p$ scattering they decrease.

\section{Search range for squarks/leptoquarks at HERA}
\begin{figure}[p]
\includegraphics{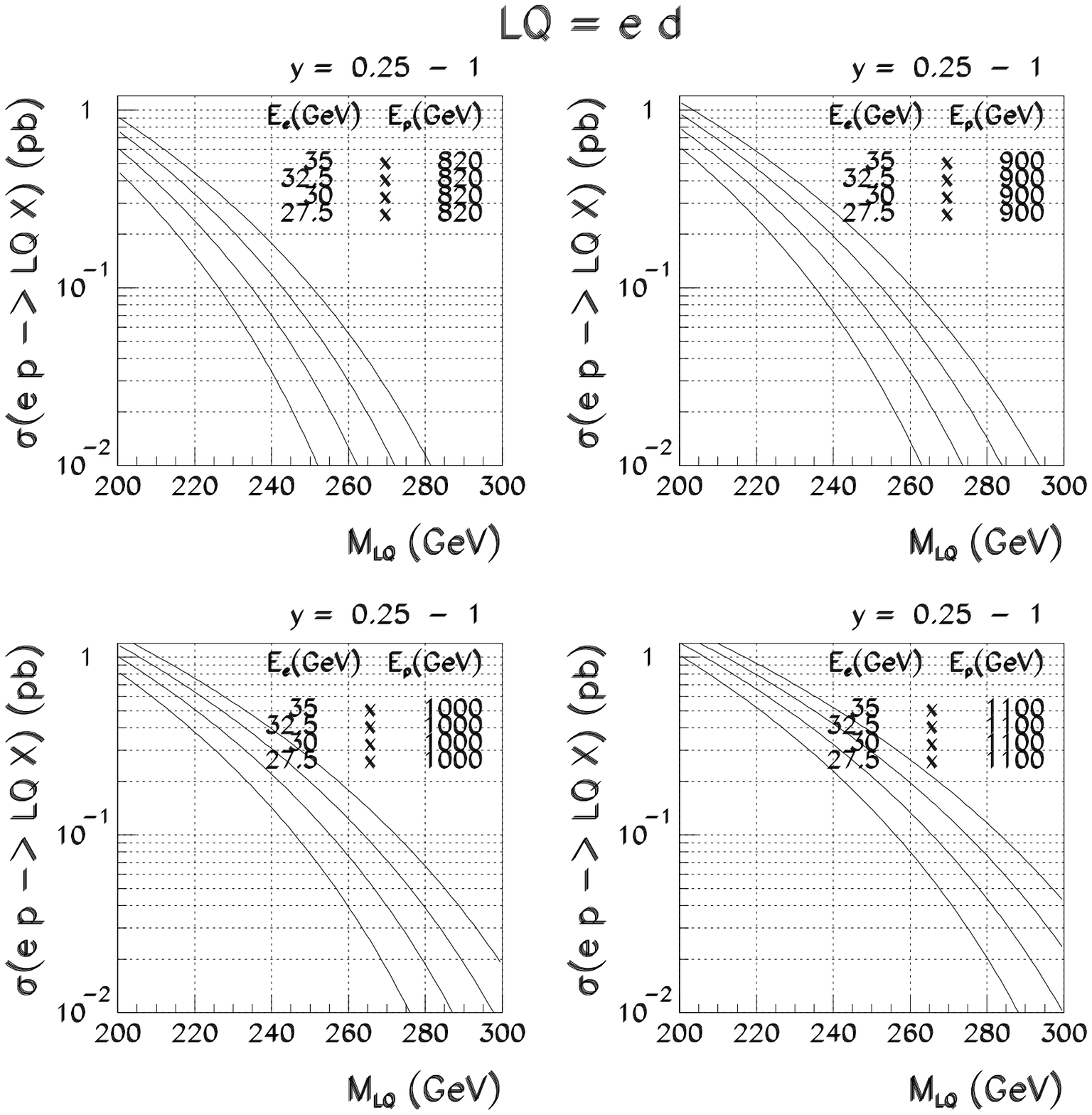}
\unitlength1cm
\begin{picture}(15,18)
\thicklines
\end{picture}
\caption{The cross section $\sigma(ep \to LQ \; X)$ for squark/leptoquark
         production by $ed$ fusion for $\lambda_{ed} = 0.045$;
         see text.}
\protect\label{f:sigloglqd}
\end{figure}

\begin{figure}[p]
\includegraphics{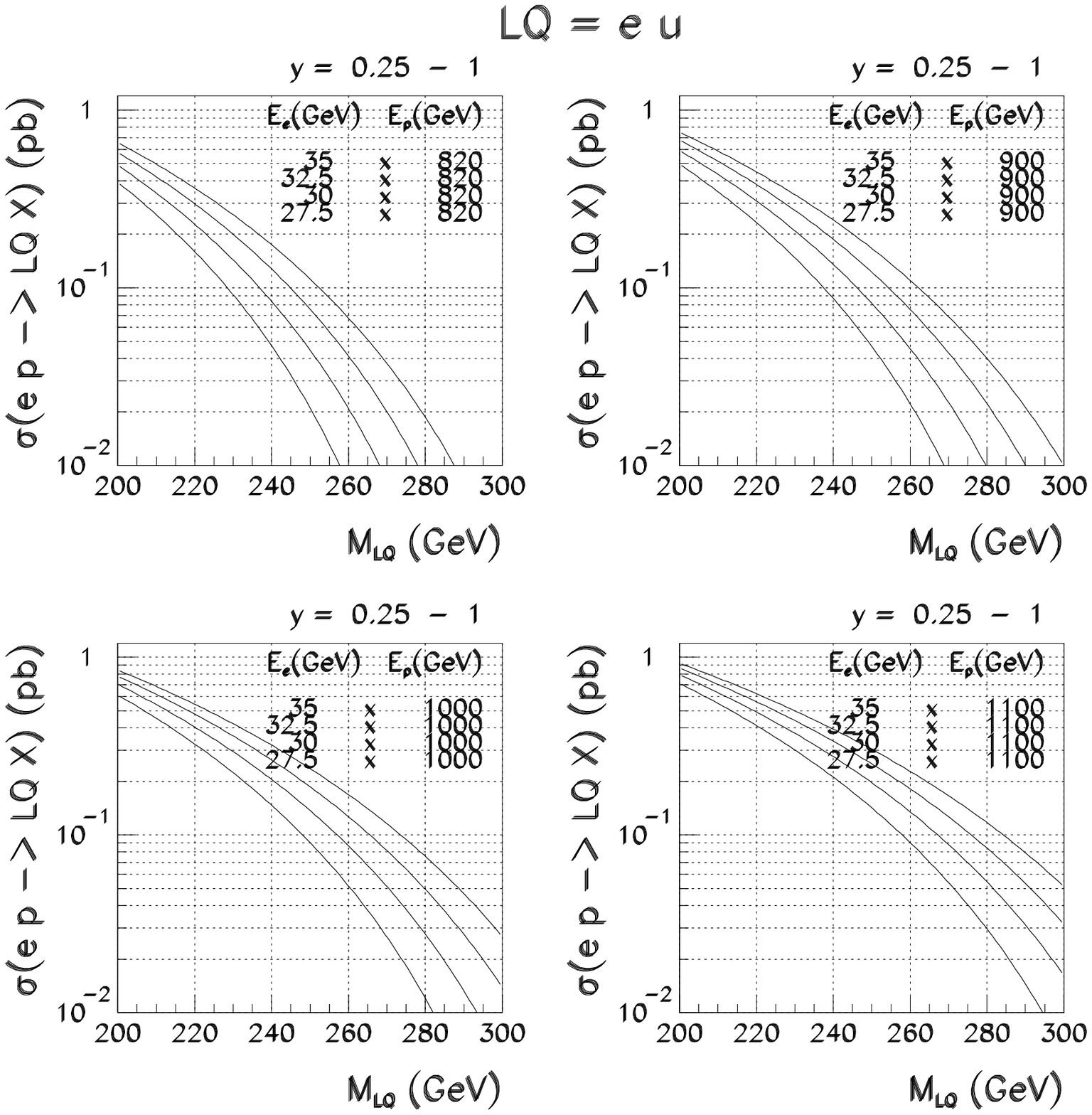}
\unitlength1cm
\begin{picture}(13.5,16)
\thicklines
\end{picture}
\caption{The cross section $\sigma(ep \to LQ \; X)$ for squark/leptoquark 
         production by $eu$ fusion for $\lambda_{eu} = 0.021$;
         see text.}
\protect\label{f:sigloglqu}
\end{figure}

\begin{figure}[p]
\includegraphics{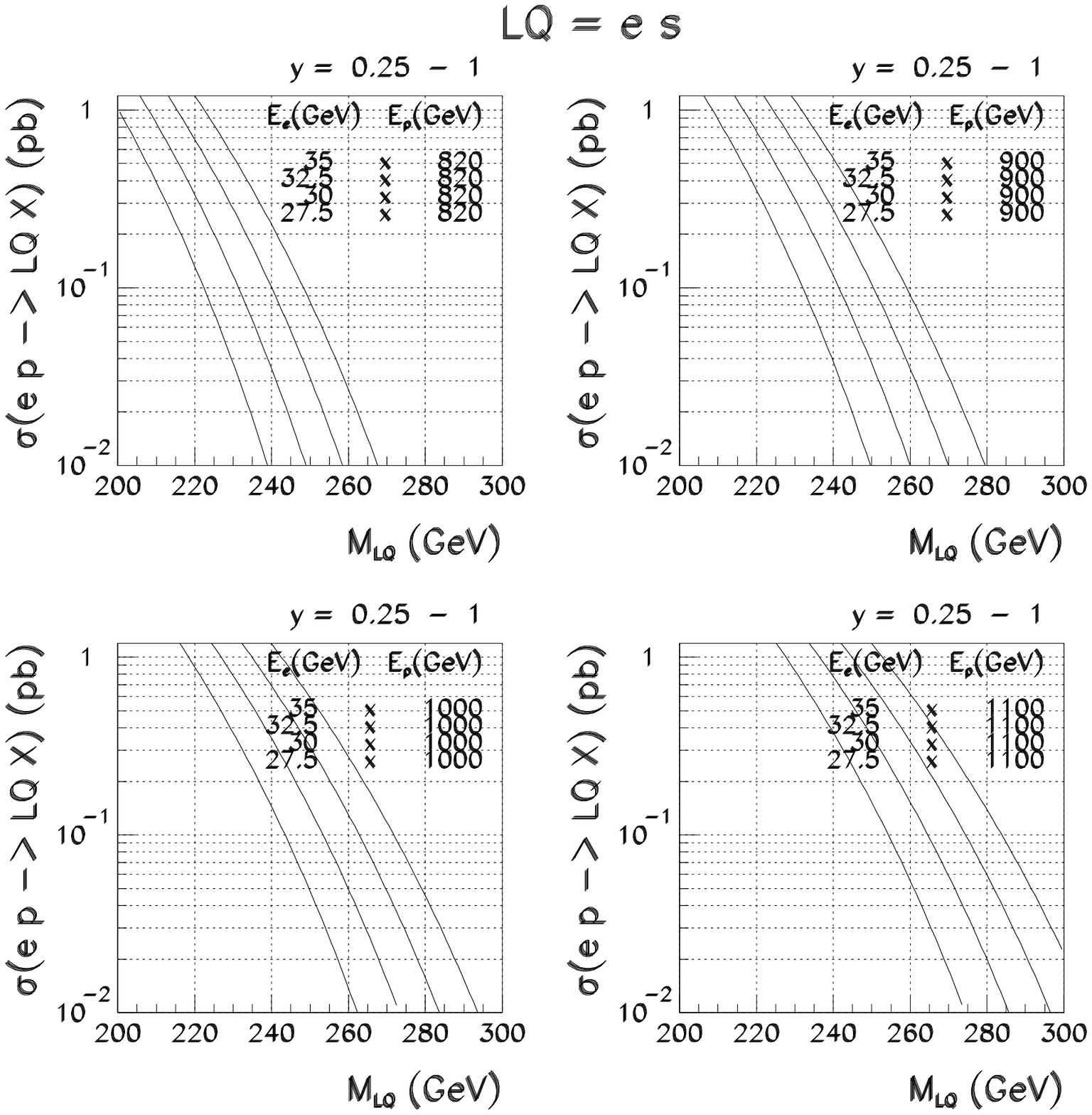}
\unitlength1cm
\begin{picture}(13.5,16)
\thicklines
\end{picture}
\caption{The cross section $\sigma(ep \to LQ \; X)$ for squark/leptoquark
         production by $es$ fusion for $\lambda_{es} = 0.7$;
         see text.}
\protect\label{f:sigloglqs}
\end{figure}

The cross sections for production of squarks/leptoquarks in $ep$ collisions by $ed$, $eu$ and $es$ fusion are shown in Figs.~\ref{f:sigloglqd},~\ref{f:sigloglqu},~\ref{f:sigloglqs} as a function of $M_{LQ}$ between 200 and 300 GeV for different combinations of beam energies. They were calculated from Eq.~\ref{lqcross}, using the MRSA parton distribution set, and integrated over $y>0.25$. As before, we assumed $\lambda_{ed} = 0.045, \lambda_{eu} = 0.021, \lambda_{es} = 0.7$. Assuming that a $\tilde{q}/LQ$ signal can be established with 10 events, a minimum cross section of 50 fb is required for an integrated luminosity of 200 pb$^{-1}$. For the nominal beam energies ($E_e = 27.5$ GeV, $E_p = 820$ GeV) this condition limits the search range to masses of 225 - 235 GeV. By raising the beam energies to (32.5 GeV, 1000 GeV) the search range is extended by about 40 GeV to 270 - 280 GeV. If the beam energies are raised even higher to (35 GeV, 1100 GeV) the search range increases to $M_{LQ} = 290 - 300$ GeV. Of course, the search range is larger (smaller) for higher (lower) values of $\lambda$.

In this study a minimum of 10 signal events were required using the LO cross section for LQ production as given by Eq.~\ref{lqcross}. QCD corrections from gluon radiation, gluon splitting and vertex corrections increase the cross section~\cite{Plehn97,Kunszt97}. However, the requirement of a minimum of 10 signal events leads basically to the same conclusions for the search range as before.

\section{\bf Conclusions}
\label{s:Conclusions}
The production rate of squarks/leptoquarks with a mass around 200 GeV can be increased substantially by raising the beam energies of HERA. The number of $LQ$ events increases 
\begin{tabbing}
\hspace{1cm}by a factor of \=1.5 for $E_e,E_p$ = \=27.5 GeV, ~900 GeV\\
                           \>2.1                 \>27.5 GeV, 1000 GeV\\
                           \>2.0                 \>30.0 GeV, ~900 GeV\\
                           \>2.7                 \>30.0 GeV, 1000 GeV\\
                           \>3.3                 \>32.5 GeV, 1000 GeV.\\
\end{tabbing}
These numbers were obtained for $M_{LQ} = 215$ GeV and represent 
averages over $e^+u$ and $e^+d$ production. For $e^+s$ leptoquark
production the gains are even larger, see Fig.~\ref{f:gains}. 

By raising the beam energies from $E_e = 27.5$ GeV, $E_p = 820$ GeV to $E_e = 32.5$ GeV, $E_p = 1000$ GeV (35 GeV, 1100 GeV) the search range for squarks/leptoquarks can be extended by about 40 (75) GeV to $M_{LQ} = 270 - 280$ $(290 - 300)$ GeV.

\section{Acknowledgements}
We thank Dr. U. Katz for useful comments.

\clearpage
\newpage
%\tabcolsep 1.2pt

%-----------------------------------------------------------------------------
\begin{table}[p]
\begin{center}
\vspace{1.5cm} 
\begin{tabular}{|ccc|c|rrr|rr|rr|}
\hline
$E_e$ & $E_p$ & $M_{LQ}$ & $x_{LQ}$ & 
\multicolumn{3}{c|}{$e^+q_i \rightarrow LQ$} &
\multicolumn{4}{c|}{$ep \rightarrow e(\nu)X$} \\
\cline{5-11}
(GeV) & (GeV) & (GeV) & & 
$N^d_{LQ}$ & $N^u_{LQ}$ & $N^s_{LQ}$ & $N^+_{NC}$ & $N^+_{CC}$ &
                                       $N^-_{NC}$ & $N^-_{CC}$ \\
\hline
\hline
27.5 &  820 & 200 & 0.44 & 10.0 & 10.0 & 10.0 & 20.1 &  4.4 & 39.0 & 53.7 \\
27.5 &  820 & 210 & 0.49 & 10.0 & 10.0 & 10.0 & 11.8 &  2.5 & 23.5 & 33.9 \\
27.5 &  820 & 220 & 0.54 & 10.0 & 10.0 & 10.0 &  6.7 &  1.3 & 13.5 & 20.3 \\
27.5 &  820 & 230 & 0.59 & 10.0 & 10.0 & 10.0 &  3.6 &  0.6 &  7.3 & 11.4 \\
\hline
27.5 &  900 & 200 & 0.40 & 13.7 & 12.7 & 18.8 & 28.7 &  7.0 & 55.2 & 75.1 \\
27.5 &  900 & 210 & 0.45 & 14.8 & 13.5 & 21.3 & 17.8 &  4.1 & 35.1 & 50.1 \\
27.5 &  900 & 220 & 0.49 & 16.1 & 14.6 & 24.7 & 10.7 &  2.3 & 21.5 & 32.2 \\
27.5 &  900 & 230 & 0.53 & 18.0 & 16.0 & 30.0 &  6.2 &  1.3 & 12.7 & 19.7 \\
\hline
27.5 & 1000 & 200 & 0.36 & 18.5 & 15.9 & 34.6 & 41.1 & 11.3 & 78.1 &104.6 \\
27.5 & 1000 & 210 & 0.40 & 21.2 & 17.8 & 43.5 & 26.6 &  7.0 & 51.9 & 73.2 \\
27.5 & 1000 & 220 & 0.44 & 24.9 & 20.4 & 57.6 & 16.9 &  4.2 & 33.8 & 49.9 \\
27.5 & 1000 & 230 & 0.48 & 30.5 & 24.3 & 82.0 & 10.5 &  2.5 & 21.4 & 32.8 \\
\hline
27.5 & 1100 & 200 & 0.33 & 23.1 & 18.6 & 55.4 & 55.1 & 16.7 &103.3 &136.3 \\
27.5 & 1100 & 210 & 0.36 & 27.7 & 21.6 & 75.3 & 36.8 & 10.7 & 71.0 & 98.7 \\
27.5 & 1100 & 220 & 0.40 & 34.3 & 26.0 &109.3 & 24.3 &  6.7 & 48.1 & 70.0 \\
27.5 & 1100 & 230 & 0.44 & 44.9 & 32.7 &173.4 & 15.8 &  4.1 & 31.9 & 48.4 \\
\hline
\hline
30.0 &  820 & 200 & 0.41 & 13.5 & 12.5 & 18.1 & 28.1 &  6.8 & 54.1 & 73.6 \\
30.0 &  820 & 210 & 0.45 & 14.4 & 13.3 & 20.3 & 17.4 &  4.0 & 34.2 & 48.9 \\
30.0 &  820 & 220 & 0.49 & 15.7 & 14.2 & 23.4 & 10.4 &  2.3 & 20.9 & 31.3 \\
30.0 &  820 & 230 & 0.54 & 17.4 & 15.6 & 28.1 &  6.0 &  1.2 & 12.3 & 19.1 \\
\hline
30.0 &  900 & 200 & 0.37 & 17.6 & 15.3 & 31.3 & 38.7 & 10.4 & 73.7 & 99.0 \\
30.0 &  900 & 210 & 0.41 & 20.0 & 17.0 & 38.8 & 24.9 &  6.4 & 48.7 & 68.8 \\
30.0 &  900 & 220 & 0.45 & 23.2 & 19.3 & 50.3 & 15.7 &  3.8 & 31.4 & 46.4 \\
30.0 &  900 & 230 & 0.49 & 28.1 & 22.7 & 69.8 &  9.7 &  2.2 & 19.6 & 30.2 \\
\hline
30.0 & 1000 & 200 & 0.33 & 22.7 & 18.4 & 53.3 & 53.8 & 16.2 &100.9 &133.3 \\
30.0 & 1000 & 210 & 0.37 & 27.1 & 21.3 & 72.0 & 35.8 & 10.3 & 69.2 & 96.3 \\
30.0 & 1000 & 220 & 0.40 & 33.5 & 25.5 &103.8 & 23.6 &  6.5 & 46.7 & 68.1 \\
30.0 & 1000 & 230 & 0.44 & 43.5 & 31.9 &163.3 & 15.3 &  4.0 & 30.9 & 46.9 \\
\hline
30.0 & 1100 & 200 & 0.30 & 27.5 & 21.0 & 80.7 & 70.8 & 23.4 &130.7 &169.7 \\
30.0 & 1100 & 210 & 0.33 & 34.0 & 25.1 &116.2 & 48.3 & 15.4 & 91.9 &126.1 \\
30.0 & 1100 & 220 & 0.37 & 43.9 & 31.2 &180.6 & 32.8 & 10.0 & 64.1 & 92.1 \\
30.0 & 1100 & 230 & 0.40 & 60.2 & 40.8 &310.4 & 22.0 &  6.4 & 44.0 & 66.0 \\
\hline
\end{tabular}
\end{center}
\vspace{-0.3cm}
\caption{The expected number of events $N^i_{LQ}$ for
the squark/leptoquark production processes $e^+q_i \rightarrow LQ$, $q_i = d,u,s$.
The number of background events from $e^{\pm}p$ NC and CC scattering
are listed in the columns $N^{\pm}_{NC}$ and $N^{\pm}_{CC}$.
The event yields are given 
in the bin
$x_{LQ} \pm 0.08$, $y > 0.25$
for an integrated luminosity of $L = 50$ pb$^{-1}$.
}
\label{t:tab1}
\end{table}

%-----------------------------------------------------------------------------
\begin{table}[p]
\begin{center}
\vspace{1.5cm} 
\begin{tabular}{|ccc|c|rrr|rr|rr|}
\hline
$E_e$ & $E_p$ & $M_{LQ}$ & $x_{LQ}$ & 
\multicolumn{3}{c|}{$e^+q_i \rightarrow LQ$} &
\multicolumn{4}{c|}{$ep \rightarrow e(\nu)X$} \\
\cline{5-11}
(GeV) & (GeV) & (GeV) & & 
$N^d_{LQ}$ & $N^u_{LQ}$ & $N^s_{LQ}$ & $N^+_{NC}$ & $N^+_{CC}$ &
                                       $N^-_{NC}$ & $N^-_{CC}$ \\
\hline
\hline
32.5 &  820 & 200 & 0.38 & 17.0 & 14.9 & 29.2 & 37.1 &  9.8 & 70.7 & 95.2 \\
32.5 &  820 & 210 & 0.41 & 19.1 & 16.5 & 35.6 & 23.7 &  6.0 & 46.5 & 65.8 \\
32.5 &  820 & 220 & 0.45 & 22.1 & 18.6 & 45.6 & 14.9 &  3.6 & 29.8 & 44.1 \\
32.5 &  820 & 230 & 0.50 & 26.4 & 21.7 & 62.1 &  9.1 &  2.0 & 18.5 & 28.5 \\
\hline
32.5 &  900 & 200 & 0.34 & 21.5 & 17.7 & 47.3 & 49.8 & 14.6 & 93.9 &124.5 \\
32.5 &  900 & 210 & 0.38 & 25.3 & 20.3 & 62.6 & 33.0 &  9.2 & 63.8 & 89.2 \\
32.5 &  900 & 220 & 0.41 & 30.9 & 24.0 & 88.2 & 21.5 &  5.7 & 42.7 & 62.4 \\
32.5 &  900 & 230 & 0.45 & 39.5 & 29.6 &135.1 & 13.8 &  3.5 & 27.9 & 42.5 \\
\hline
32.5 & 1000 & 200 & 0.31 & 26.7 & 20.6 & 75.8 & 67.9 & 22.1 &125.6 &163.5 \\
32.5 & 1000 & 210 & 0.34 & 32.9 & 24.5 &108.1 & 46.1 & 14.4 & 88.0 &121.0 \\
32.5 & 1000 & 220 & 0.37 & 42.2 & 30.3 &166.2 & 31.2 &  9.3 & 61.0 & 88.0 \\
32.5 & 1000 & 230 & 0.41 & 57.3 & 39.3 &281.9 & 20.8 &  5.9 & 41.7 & 62.7 \\
\hline
32.5 & 1100 & 200 & 0.28 & 31.5 & 23.0 &109.9 & 88.1 & 31.3 &159.9 &204.5 \\
32.5 & 1100 & 210 & 0.31 & 40.0 & 28.1 &165.5 & 61.0 & 21.0 &114.4 &154.8 \\
32.5 & 1100 & 220 & 0.34 & 53.3 & 35.9 &271.0 & 42.2 & 13.9 & 81.4 &115.7 \\
32.5 & 1100 & 230 & 0.37 & 75.6 & 48.3 &494.8 & 29.0 &  9.2 & 57.4 & 85.2 \\
\hline
\hline
35.0 &  820 & 200 & 0.35 & 20.6 & 17.1 & 43.1 & 47.0 & 13.5 & 88.8 &118.2 \\
35.0 &  820 & 210 & 0.38 & 24.0 & 19.5 & 56.2 & 30.9 &  8.5 & 60.0 & 84.1 \\
35.0 &  820 & 220 & 0.42 & 29.0 & 22.9 & 77.8 & 20.0 &  5.2 & 39.8 & 58.4 \\
35.0 &  820 & 230 & 0.46 & 36.6 & 27.9 &116.6 & 12.7 &  3.1 & 25.8 & 39.4 \\
\hline
35.0 &  900 & 200 & 0.32 & 25.1 & 19.7 & 66.4 & 62.1 & 19.6 &115.5 &151.3 \\
35.0 &  900 & 210 & 0.35 & 30.6 & 23.3 & 92.8 & 41.9 & 12.7 & 80.3 &111.0 \\
35.0 &  900 & 220 & 0.38 & 38.7 & 28.4 &139.2 & 28.0 &  8.1 & 55.2 & 79.9 \\
35.0 &  900 & 230 & 0.42 & 51.8 & 36.4 &229.8 & 18.5 &  5.1 & 37.3 & 56.2 \\
\hline
35.0 & 1000 & 200 & 0.29 & 30.4 & 22.5 &101.6 & 83.2 & 29.0 &151.7 &194.9 \\
35.0 & 1000 & 210 & 0.32 & 38.4 & 27.4 &151.3 & 57.4 & 19.3 &108.1 &146.8 \\
35.0 & 1000 & 220 & 0.35 & 50.8 & 34.7 &244.5 & 39.5 & 12.8 & 76.6 &109.1 \\
35.0 & 1000 & 230 & 0.38 & 71.4 & 46.3 &439.8 & 27.0 &  8.3 & 53.7 & 79.8 \\
\hline
35.0 & 1100 & 200 & 0.26 & 35.1 & 24.7 &142.2 &107.1 & 40.5 &191.0 &240.4 \\
35.0 & 1100 & 210 & 0.29 & 45.6 & 30.8 &222.1 & 74.9 & 27.6 &138.4 &184.7 \\
35.0 & 1100 & 220 & 0.31 & 62.3 & 40.0 &378.9 & 52.5 & 18.7 &100.1 &140.5 \\
35.0 & 1100 & 230 & 0.34 & 90.9 & 55.2 &725.7 & 36.8 & 12.5 & 72.0 &105.5 \\
\hline
\end{tabular}
\end{center}
\vspace{-0.3cm}
\caption{Continuation of Table 1.}
\label{t:tab2}
\end{table}

%-----------------------------------------------------------------------------
\begin{table}[p]
\begin{center}
\vspace{1.5cm} 
\begin{tabular}{|ccc|c|rrr|rrr|rrr|}
\hline
$E_e$ & $E_p$ & $M_{LQ}$ & $x_{LQ}$ & 
\multicolumn{3}{c|}{$e^+d \rightarrow LQ$} &
\multicolumn{3}{c|}{$e^+u \rightarrow LQ$} &
\multicolumn{3}{c|}{$e^+s \rightarrow LQ$} \\
\cline{5-13}
(GeV) & (GeV) & (GeV) & &
$G$ & $S^+_{NC}$ & $S^+_{CC}$ &
$G$ & $S^+_{NC}$ & $S^+_{CC}$ &
$G$ & $S^+_{NC}$ & $S^+_{CC}$ \\
\hline
\hline
27.5 &  820 & 200 & 0.44 & 1.0 & 0.5 & 2.3 & 1.0 & 0.5 & 2.3 & 1.0 & 0.5 & 2.3 \\
27.5 &  820 & 210 & 0.49 & 1.0 & 0.8 & 4.1 & 1.0 & 0.8 & 4.1 & 1.0 & 0.8 & 4.1 \\
27.5 &  820 & 220 & 0.54 & 1.0 & 1.5 & 7.7 & 1.0 & 1.5 & 7.7 & 1.0 & 1.5 & 7.7 \\
27.5 &  820 & 230 & 0.59 & 1.0 & 2.8 &15.5 & 1.0 & 2.8 &15.5 & 1.0 & 2.8 &15.5 \\
\hline
27.5 &  900 & 200 & 0.40 & 1.4 & 0.5 & 2.0 & 1.3 & 0.4 & 1.8 & 1.9 & 0.7 & 2.7 \\
27.5 &  900 & 210 & 0.45 & 1.5 & 0.8 & 3.6 & 1.4 & 0.8 & 3.3 & 2.1 & 1.2 & 5.1 \\
27.5 &  900 & 220 & 0.49 & 1.6 & 1.5 & 6.9 & 1.5 & 1.4 & 6.2 & 2.5 & 2.3 &10.5 \\
27.5 &  900 & 230 & 0.53 & 1.8 & 2.9 &14.1 & 1.6 & 2.6 &12.6 & 3.0 & 4.8 &23.5 \\
\hline
27.5 & 1000 & 200 & 0.36 & 1.9 & 0.5 & 1.6 & 1.6 & 0.4 & 1.4 & 3.5 & 0.8 & 3.1 \\
27.5 & 1000 & 210 & 0.40 & 2.1 & 0.8 & 3.0 & 1.8 & 0.7 & 2.5 & 4.4 & 1.6 & 6.2 \\
27.5 & 1000 & 220 & 0.44 & 2.5 & 1.5 & 5.9 & 2.0 & 1.2 & 4.8 & 5.8 & 3.4 &13.7 \\
27.5 & 1000 & 230 & 0.48 & 3.1 & 2.9 &12.4 & 2.4 & 2.3 & 9.8 & 8.2 & 7.8 &33.3 \\
\hline
27.5 & 1100 & 200 & 0.33 & 2.3 & 0.4 & 1.4 & 1.9 & 0.3 & 1.1 & 5.5 & 1.0 & 3.3 \\
27.5 & 1100 & 210 & 0.36 & 2.8 & 0.8 & 2.6 & 2.2 & 0.6 & 2.0 & 7.5 & 2.0 & 7.0 \\
27.5 & 1100 & 220 & 0.40 & 3.4 & 1.4 & 5.1 & 2.6 & 1.1 & 3.9 &10.9 & 4.5 &16.2 \\
27.5 & 1100 & 230 & 0.44 & 4.5 & 2.8 &10.8 & 3.3 & 2.1 & 7.9 &17.3 &11.0 &41.8 \\
\hline
\hline
30.0 &  820 & 200 & 0.41 & 1.3 & 0.5 & 2.0 & 1.3 & 0.4 & 1.8 & 1.8 & 0.6 & 2.7 \\
30.0 &  820 & 210 & 0.45 & 1.4 & 0.8 & 3.6 & 1.3 & 0.8 & 3.3 & 2.0 & 1.2 & 5.1 \\
30.0 &  820 & 220 & 0.49 & 1.6 & 1.5 & 6.9 & 1.4 & 1.4 & 6.3 & 2.3 & 2.2 &10.4 \\
30.0 &  820 & 230 & 0.54 & 1.7 & 2.9 &14.2 & 1.6 & 2.6 &12.7 & 2.8 & 4.7 &23.0 \\
\hline
30.0 &  900 & 200 & 0.37 & 1.8 & 0.5 & 1.7 & 1.5 & 0.4 & 1.5 & 3.1 & 0.8 & 3.0 \\
30.0 &  900 & 210 & 0.41 & 2.0 & 0.8 & 3.1 & 1.7 & 0.7 & 2.7 & 3.9 & 1.6 & 6.1 \\
30.0 &  900 & 220 & 0.45 & 2.3 & 1.5 & 6.1 & 1.9 & 1.2 & 5.1 & 5.0 & 3.2 &13.2 \\
30.0 &  900 & 230 & 0.49 & 2.8 & 2.9 &12.7 & 2.3 & 2.4 &10.3 & 7.0 & 7.2 &31.6 \\
\hline
30.0 & 1000 & 200 & 0.33 & 2.3 & 0.4 & 1.4 & 1.8 & 0.3 & 1.1 & 5.3 & 1.0 & 3.3 \\
30.0 & 1000 & 210 & 0.37 & 2.7 & 0.8 & 2.6 & 2.1 & 0.6 & 2.1 & 7.2 & 2.0 & 7.0 \\
30.0 & 1000 & 220 & 0.40 & 3.3 & 1.4 & 5.2 & 2.6 & 1.1 & 3.9 &10.4 & 4.4 &16.0 \\
30.0 & 1000 & 230 & 0.44 & 4.4 & 2.8 &11.0 & 3.2 & 2.1 & 8.0 &16.3 &10.7 &41.1 \\
\hline
30.0 & 1100 & 200 & 0.30 & 2.7 & 0.4 & 1.2 & 2.1 & 0.3 & 0.9 & 8.1 & 1.1 & 3.5 \\
30.0 & 1100 & 210 & 0.33 & 3.4 & 0.7 & 2.2 & 2.5 & 0.5 & 1.6 &11.6 & 2.4 & 7.6 \\
30.0 & 1100 & 220 & 0.37 & 4.4 & 1.3 & 4.4 & 3.1 & 1.0 & 3.1 &18.1 & 5.5 &18.1 \\
30.0 & 1100 & 230 & 0.40 & 6.0 & 2.7 & 9.4 & 4.1 & 1.9 & 6.4 &31.0 &14.1 &48.7 \\
\hline
\end{tabular}
\end{center}
\vspace{-0.3cm}
\caption{
The gain $G = N_{LQ}(E_e,E_p,M_{LQ})/N_{LQ}(27.5,820,M_{LQ})$ 
and the signal to background ratios
$S^+_{NC,CC} = N_{LQ}/N^+_{NC,CC}$
for the processes $e^+d \rightarrow LQ$, 
$e^+u \rightarrow LQ$ and
$e^+s \rightarrow LQ$. 
}
\label{t:tab3}
\end{table}

%-----------------------------------------------------------------------------
\begin{table}[p]
\begin{center}
\vspace{1.5cm} 
\begin{tabular}{|ccc|c|rrr|rrr|rrr|}
\hline
$E_e$ & $E_p$ & $M_{LQ}$ & $x_{LQ}$ & 
\multicolumn{3}{c|}{$e^+d \rightarrow LQ$} &
\multicolumn{3}{c|}{$e^+u \rightarrow LQ$} &
\multicolumn{3}{c|}{$e^+s \rightarrow LQ$} \\
\cline{5-13}
(GeV) & (GeV) & (GeV) & &
$G$ & $S^+_{NC}$ & $S^+_{CC}$ &
$G$ & $S^+_{NC}$ & $S^+_{CC}$ &
$G$ & $S^+_{NC}$ & $S^+_{CC}$ \\
\hline
\hline
32.5 &  820 & 200 & 0.38 & 1.7 & 0.5 & 1.7 & 1.5 & 0.4 & 1.5 & 2.9 & 0.8 & 3.0 \\
32.5 &  820 & 210 & 0.41 & 1.9 & 0.8 & 3.2 & 1.6 & 0.7 & 2.7 & 3.6 & 1.5 & 5.9 \\
32.5 &  820 & 220 & 0.45 & 2.2 & 1.5 & 6.2 & 1.9 & 1.2 & 5.2 & 4.6 & 3.1 &12.8 \\
32.5 &  820 & 230 & 0.50 & 2.6 & 2.9 &12.9 & 2.2 & 2.4 &10.6 & 6.2 & 6.8 &30.4 \\
\hline
32.5 &  900 & 200 & 0.34 & 2.1 & 0.4 & 1.5 & 1.8 & 0.4 & 1.2 & 4.7 & 0.9 & 3.2 \\
32.5 &  900 & 210 & 0.38 & 2.5 & 0.8 & 2.7 & 2.0 & 0.6 & 2.2 & 6.3 & 1.9 & 6.8 \\
32.5 &  900 & 220 & 0.41 & 3.1 & 1.4 & 5.4 & 2.4 & 1.1 & 4.2 & 8.8 & 4.1 &15.4 \\
32.5 &  900 & 230 & 0.45 & 4.0 & 2.9 &11.4 & 3.0 & 2.1 & 8.5 &13.5 & 9.8 &38.9 \\
\hline
32.5 & 1000 & 200 & 0.31 & 2.7 & 0.4 & 1.2 & 2.1 & 0.3 & 0.9 & 7.6 & 1.1 & 3.4 \\
32.5 & 1000 & 210 & 0.34 & 3.3 & 0.7 & 2.3 & 2.5 & 0.5 & 1.7 &10.8 & 2.3 & 7.5 \\
32.5 & 1000 & 220 & 0.37 & 4.2 & 1.4 & 4.5 & 3.0 & 1.0 & 3.2 &16.6 & 5.3 &17.8 \\
32.5 & 1000 & 230 & 0.41 & 5.7 & 2.8 & 9.7 & 3.9 & 1.9 & 6.6 &28.2 &13.5 &47.6 \\
\hline
32.5 & 1100 & 200 & 0.28 & 3.1 & 0.4 & 1.0 & 2.3 & 0.3 & 0.7 &11.0 & 1.2 & 3.5 \\
32.5 & 1100 & 210 & 0.31 & 4.0 & 0.7 & 1.9 & 2.8 & 0.5 & 1.3 &16.6 & 2.7 & 7.9 \\
32.5 & 1100 & 220 & 0.34 & 5.3 & 1.3 & 3.8 & 3.6 & 0.9 & 2.6 &27.1 & 6.4 &19.4 \\
32.5 & 1100 & 230 & 0.37 & 7.6 & 2.6 & 8.3 & 4.8 & 1.7 & 5.3 &49.5 &17.1 &54.0 \\
\hline
\hline
35.0 &  820 & 200 & 0.35 & 2.1 & 0.4 & 1.5 & 1.7 & 0.4 & 1.3 & 4.3 & 0.9 & 3.2 \\
35.0 &  820 & 210 & 0.38 & 2.4 & 0.8 & 2.8 & 2.0 & 0.6 & 2.3 & 5.6 & 1.8 & 6.6 \\
35.0 &  820 & 220 & 0.42 & 2.9 & 1.4 & 5.5 & 2.3 & 1.1 & 4.4 & 7.8 & 3.9 &14.9 \\
35.0 &  820 & 230 & 0.46 & 3.7 & 2.9 &11.7 & 2.8 & 2.2 & 8.9 &11.7 & 9.2 &37.2 \\
\hline
35.0 &  900 & 200 & 0.32 & 2.5 & 0.4 & 1.3 & 2.0 & 0.3 & 1.0 & 6.6 & 1.1 & 3.4 \\
35.0 &  900 & 210 & 0.35 & 3.1 & 0.7 & 2.4 & 2.3 & 0.6 & 1.8 & 9.3 & 2.2 & 7.3 \\
35.0 &  900 & 220 & 0.38 & 3.9 & 1.4 & 4.8 & 2.8 & 1.0 & 3.5 &13.9 & 5.0 &17.2 \\
35.0 &  900 & 230 & 0.42 & 5.2 & 2.8 &10.2 & 3.6 & 2.0 & 7.2 &23.0 &12.4 &45.1 \\
\hline
35.0 & 1000 & 200 & 0.29 & 3.0 & 0.4 & 1.0 & 2.2 & 0.3 & 0.8 &10.2 & 1.2 & 3.5 \\
35.0 & 1000 & 210 & 0.32 & 3.8 & 0.7 & 2.0 & 2.7 & 0.5 & 1.4 &15.1 & 2.6 & 7.8 \\
35.0 & 1000 & 220 & 0.35 & 5.1 & 1.3 & 4.0 & 3.5 & 0.9 & 2.7 &24.4 & 6.2 &19.1 \\
35.0 & 1000 & 230 & 0.38 & 7.1 & 2.6 & 8.6 & 4.6 & 1.7 & 5.6 &44.0 &16.3 &52.7 \\
\hline
35.0 & 1100 & 200 & 0.26 & 3.5 & 0.3 & 0.9 & 2.5 & 0.2 & 0.6 &14.2 & 1.3 & 3.5 \\
35.0 & 1100 & 210 & 0.29 & 4.6 & 0.6 & 1.7 & 3.1 & 0.4 & 1.1 &22.2 & 3.0 & 8.1 \\
35.0 & 1100 & 220 & 0.31 & 6.2 & 1.2 & 3.3 & 4.0 & 0.8 & 2.1 &37.9 & 7.2 &20.3 \\
35.0 & 1100 & 230 & 0.34 & 9.1 & 2.5 & 7.2 & 5.5 & 1.5 & 4.4 &72.6 &19.7 &57.9 \\
\hline
\end{tabular}
\end{center}
\vspace{-0.3cm}
\caption{Continuation of table 3.}
\label{t:tab4}
\end{table}

%-----------------------------------------------------------------------------
%figure 1 

%figure 2

%figure 3

%figure 4

%figure 5

%figure 6

%figure 7

%figure 8

%figure 9

%figure 10

%figure 11

%figure 12
\end{document}